\documentclass[%
 aip,
cp,  % Conference Proceedings
 amsmath,amssymb,%nobibnotes,
 reprint,%
]{revtex4-2}

\usepackage{graphicx}% Include figure files
\usepackage{dcolumn}% Align table columns on decimal point
\usepackage{bm}% bold math
\usepackage{braket}
\usepackage[utf8]{inputenc}
\usepackage[T1]{fontenc}
\usepackage{mathptmx} 

\begin{document}

\title{Chaos, Phase Transitions and Curvature Invariants of (Rotating, Warped, Massive) BTZ Black Holes }% Force line breaks with \\

\author{Mahdis Ghodrati} % Write as First name Surname
 \email[]{mahdis.ghodrati@apctp.org}
\affiliation{
Asia Pacific Center for Theoretical Physics, Pohang 37673 KOREA % Force line breaks with \\ if necessary
}
\date{\today} % It is always \today, today, but any date may be explicitly specified
              % Not printed for conference proceedings

\begin{abstract}
Combining several results from the previous works of the author, three main subtleties will be clarified here. First, similar to the rotating BTZ black holes, we show how for the warped BTZ black holes, the consideration of the \textbf{effective} temperatures could solve the seemingly unsaturation issue of the chaos bound. Second, comparing the Hawking-Page phase diagrams of BTZ and warped BTZ black holes in the topologically massive gravity and new massive gravity theories, we show how the characteristics of the action would specify the behaviors of the chaos modes, and there we emphasize the importance of using the local and "physical" thermodynamical ensembles for studying the boundary chaos modes in warped CFTs and also the connections with the bulk reconstruction. Third, we propose that the boundary modular scrambling modes which saturate the modular chaos bound are related to the curvature invariants in the geometrical side.
\end{abstract}

\maketitle

\section{\label{sec:level1} INTRODUCTION}
In holography and gauge/gravity duality, various examples and tools have been introduced which can depict how the properties of the bulk geometry side are related to the pattern of entanglement and correlations among the degrees of freedom in the boundary field theory side, connecting boundary algebra to bulk geometrical structures. The other motivation of constructing such tools, is to explicitly show how the dynamics in the bulk is related to the dynamics in the boundary side. In fact, in the AdS case, it has been shown that gravity even knows about the effective form of \textbf{storage} and \textbf{processing} of the quantum information in the strongly coupled and chaotic conformal field theories (CFTs).  As an example between these algebra in the CFT side and gravity side, it has been shown that, the probes of the quantum chaos (OTOC) can compute the classical gravitational scattering processes near the black holes and there the maximum Lyapunov exponent is controlled by the gravitational redshift. Here, we examine three cases in the gauge/gravity duality and try to make clear several subtle points, which helps to get a better and more unified picture of the bulk reconstruction from the boundary algebra and chaotic modes.

\section{Chaos in holography}
In \cite{Maldacena:2015waa}, the OTOC for chaotic system which can be written as
\begin{gather}
F(t, \vec{x} ) = \langle V(0) W(t, \vec{x} ) V(0) W (t, \vec{x} ) \rangle,
\end{gather}
has been calculated and there it was shown that it vanishes at the later times. It also has been shown that for "holographic" systems, it decays exponentially as
\begin{gather}
\frac{\langle V(0) W(t, \vec{x} ) V(0) W (t, \vec{x} ) \rangle  }{ \langle V(0) V(0 ) \rangle \langle W(t , \vec{x} ) W (t, \vec{x} ) \rangle }= 1- \epsilon_{\Delta_V \Delta_W} \text{exp}  \left \lbrack \lambda_L \left ( t - t_* - \frac{|\vec{x} |}{v_B} \right ) \right \rbrack,
\end{gather}
where $\epsilon_{\Delta_V \Delta_W} $ contains information about the operators $V$ and $W$, $\Lambda_L$ is the Lyapunov exponent, and $v_B$ is the butterfly velocity. A bound on the Lyapunov exponent in large-N systems has been found as $\lambda_L  \le \frac{2\pi}{\beta}$, where the black holes saturate it. In addition, in \cite{Maldacena:2015waa}, it has been shown that the Sachdev-Ye-Kitaev (SYK) models also saturate this bound. Interestingly, in \cite{DeBoer:2019kdj}, it was shown that the algebra of modular chaos scrambling modes in the boundary would directly lead to the Riemannian curvature in the bulk; check also \cite{Chen:2021lnq, Kibe:2021gtw, Ghodrati:2020vzm, Ghodrati:2021ozc} and the references therein. There are however a few subtle points in these connections which should be clarified to get a correct picture, which we discuss in the next section.

\subsection{Chaos in rotating BTZ black holes}
\begin{figure}[ht!]
\includegraphics[width=0.35\textwidth]{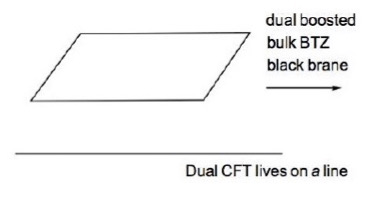}% Here is how to import EPS art
\caption{\label{fig:BTZ} The CFT on a spatial line is dual to a boosted BTZ black brane.}
\end{figure}

For the $2d$ CFTs, the effective temperatures of the left and right moving modes, i.e, $T_+$ and $T_-$ are equal. The picture for the correspondence is that the dual bulk geometry of a boosted BTZ black brane is the CFT on a spatial line shown in figure \ref{fig:BTZ}.

\begin{figure}[ht!]
\includegraphics[width=0.3\textwidth]{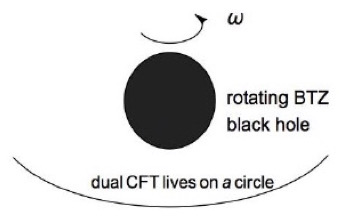}% Here is how to import EPS art
\caption{\label{fig:rotBTZ} The CFT on a circle is dual to a "rotating" BTZ black hole.}
\end{figure}

On the other hand, for the CFT on a circle, the dual bulk geometry is a rotating BTZ black hole with the schematic shown in figure \ref{fig:rotBTZ}.

The Lyapunov exponent for CFTs on the circle then holographically and using the dual rotating BTZ black holes has been found as
\begin{gather}
\lambda_L^{\pm} = \frac{2\pi}{\beta} \frac{1}{ 1 \mp \ell \Omega}.
\end{gather}

The interesting point here is that the onset of chaos is controlled by $\lambda_L^-$. So, the question is that if there is a non-saturation of the chaos bound in this example, as we have the following inequality relation
\begin{gather}
\lambda_L^- \le \frac{2\pi}{\beta} \le \lambda_L^+.
\end{gather}

This issue and its resolution have been discussed in \cite{Jahnke:2019gxr} where first the OTOCs of rotating black holes have been found using two ways. The first method is the \textbf{elastic eikonal gravity approximation} where the OTOCs have the picture of high energy shock waves, and the second method is using the Chern-Simon formulation of $3d$ gravity where the dual boundary CFT side is two copies of Schwarzian-like action. The result for the OTOC in this case would be
\begin{gather}
OTOC( t, \varphi_{12}) \approx 1+ C_1 e^{\frac{2\pi}{\beta_+} (t+ \ell \varphi_{12} )} +C_2 e^{\frac{2\pi}{\beta_-} (t- \ell \varphi_{12} )}, 
\end{gather} 
where the inverse of the "effective temperatures" are $\beta_{\pm} = \beta (1 \mp \ell \Omega)$. Considering these effective temperatures for the left and right moving modes instead of $\beta$ then solves our issue, and therefore there is no violation in this case. Note also in the string theory picture and in the full D1-D5 brane system, these are the inverse temperatures of the left and right moving excitations on the effective strings.

So using this idea of taking the "effective" temperature for the left and right moving modes, we then consider another example, which is the warped CFT case.

\subsection{Chaos in warped CFT/ warped BTZ}
Warped CFTs (WCFTs) \cite{Detournay:2012pc, Anninos:2008fx, Song:2017czq, Song:2011sr, Ghodrati:2016ggy} are $2d$ QFTs with a chiral scaling symmetry that acts only on the right moving modes $x^- \to \lambda x^-$, which is in contrast with the CFT case which has a second independent scaling symmetry $x^+ \to \bar{\lambda} x^+$. Also, these theories don't satisfy the Brown-Henneaux's boundary condition. Their extended algebra is one copy of Virasoro, and one copy of $U(1)$ Kac-Moody algebra, with the global symmetry group of $SL(2,R) \times U(1)$, instead of the CFT case which has the global $SL(2,R) \times SL(2,R)$ symmetry and two copies of Virasoro algebra. So in WCFT case, the right moving current generates a left global symmetry where these symmetries then lead to a new type of modular transformation on the torus and new thermodynamical behavior.

The black hole in the dual bulk side could be written as the deformed version of AdS metric which is either squashed or squeezed and can be formulated as
\begin{gather}
g_{WAdS} = g_{AdS_3} -2 H^2 \xi \otimes \xi,
\end{gather}
where $H^2$ is the deformation parameter and $\xi^\mu$ is constant norm Killing vector of $SL(2,R)$ isometry of $AdS_3$.
 
For these theories, in \cite{Apolo:2018oqv},  the entanglement entropy of an excited state on a single interval has been calculated. There, they showed that WCFTs are maximally chaotic which is compatible with the existence of a black hole in the holographic dual picture for these theories.

The main point is that similar to rotating case, for the WCFTs, also two "effective" temperatures based on the behavior of the left and right moving modes should be defined and then the bound on chaos should be considered using these new effective temperatures. The left and right-moving modes and temperatures have the following relations among them
\begin{gather}
x^+= e^{\frac{2\pi}{\beta} w^+}, \ \ \ \ \ \ x^-= w^-+ \left(\frac{\bar{\beta} }{\beta} - \frac{2\pi \mu}{\beta} \right) w^+. 
\end{gather}

The inverse of these effective temperatures depend on the tilt parameter $H$ or $k$, Virasoro central charge $c$, and $U(1)$ Kac-Moody charge $h$, and follow the relations 
\begin{gather}\label{eq:spectral}
-i P_0^{\text{vac}} \left ( \frac{\bar{\beta} }{\beta} - \frac{2\pi \mu}{\beta} \right )= \mu q_\psi, \ \ \ \ \ \ 
\beta=\tilde{\beta}_\psi \equiv \frac{1\pi}{\sqrt{24 \tilde{h}_\psi/c-1} }.
\end{gather}

Here $\beta$ or $\ell$ denotes the lengths of the subsystem along Virasoro direction, and $\bar{\beta}$ or $\bar{\ell}$ denotes the lengths of the subsystem along the $U(1)$ direction and $\mu$ is the spectral form factor. Also, there is the relation $\tilde{h}_\psi= h_\psi- q^2/k$ where $h_\psi$ and $q_\psi$ are the conformal weight and charge of $\ket{\psi}$ respectively. Here $\tilde{h}_i$ is the conformal weight in the Sugawara basis.

These effective temperatures can also be defined based on the following warped conformal transformation \cite{Song:2016gtd}
\begin{gather}
x^+= e^{\frac{2\pi}{\beta} w^+}, \ \ \ \ \ \ x^-=w^- + \left ( \frac{\bar{\beta} }{\beta} - \frac{2\pi \mu}{\beta} \right ) w^+.
\end{gather}

So, the tension and the temperature that the left-moving modes versus right-moving modes "feel" are different but in the coordinate system of the "observer", for each mode, the chaos bound is still saturated and similar to the BTZ case there would not be any violation or unsaturation.

\begin{figure}[ht!]
\includegraphics[width=0.5\textwidth]{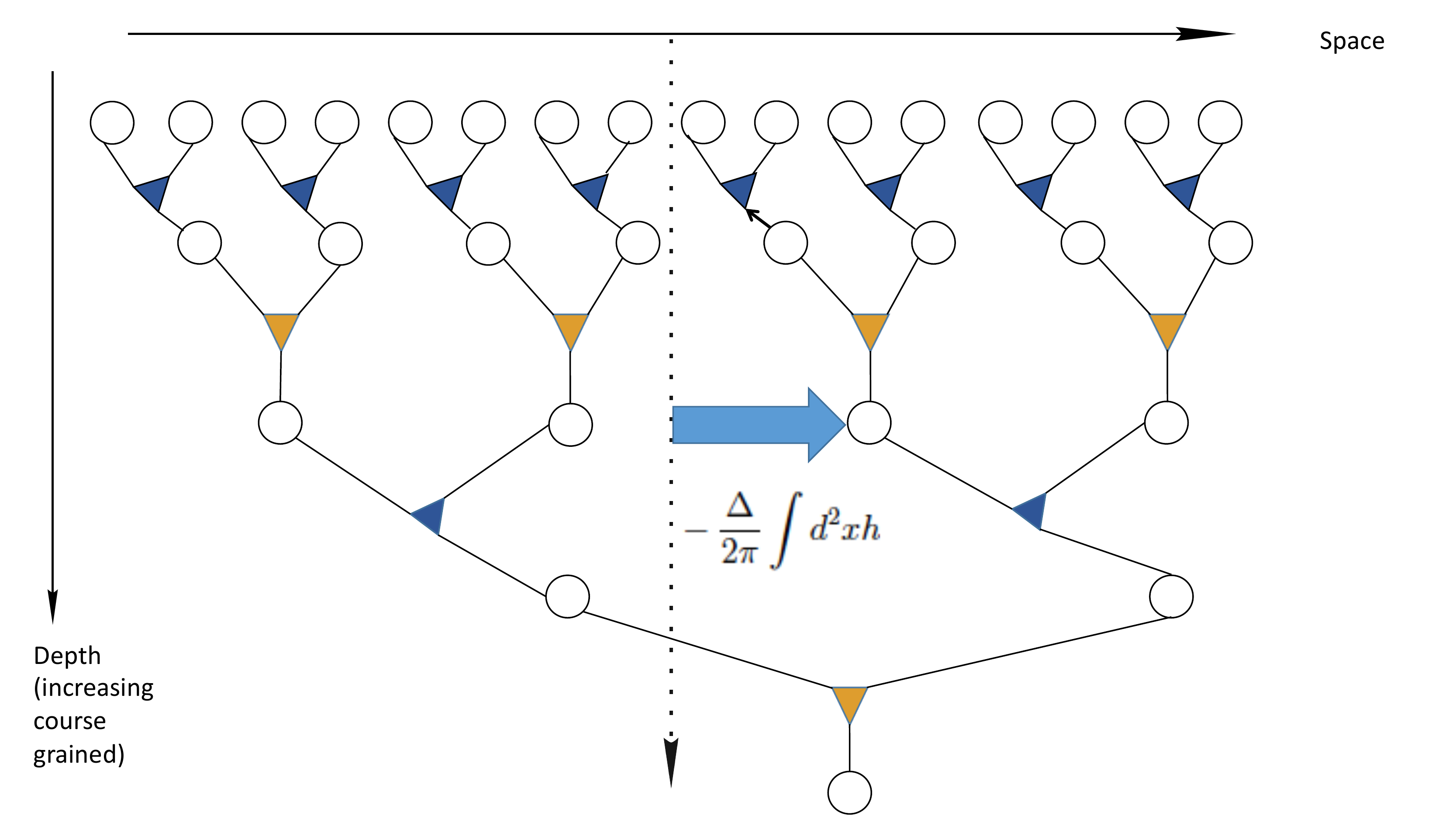}% Here is how to import EPS art
\caption{\label{fig:TNwarped} The tensor network structure of warped CFTs discussed in \cite{Ghodrati:2019bzz}. Permission is granted to reuse the figure here.}
\end{figure}

\begin{figure}[ht!]
\includegraphics[width=0.18\textwidth]{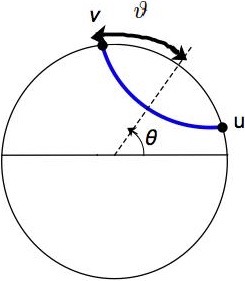} \ \ \ \ \  \ \ \ \ \ % Here is how to import EPS art 
\includegraphics[width=0.25\textwidth]{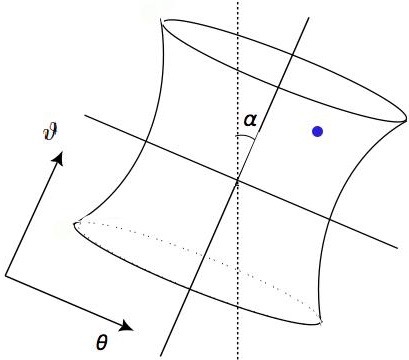}% Here is how to import EPS art
\caption{\label{fig:Tiltwarped} The tilt angle in warped CFTs, which affects kinematic space, chaos bounds and bulk reconstructions, \cite{Ghodrati:2019bzz}. Permission is granted to reuse the figure here.}
\end{figure}

This could also be noticed from the tensor network structure of the warped CFTs presented in \cite{Ghodrati:2019bzz} and shown in figure \ref{fig:TNwarped} here. The tilt angle of warped CFTs which affects the procedure of bulk reconstruction from the boundary algebra of these theories, are also shown in figure \ref{fig:Tiltwarped}.

The parameter $\mu$ used in equation \ref{eq:spectral}, which is the spectral flow parameter, is related to $\alpha$ which is the tilt angle of the cylinder. The chaos bound for the right versus left moving modes depends on this parameter.  As the result, the structure of these theories and its algebra would correspondingly determine the behavior of the boundary chaos modes, the tensor network structure, the modular flows, the effective temperature, the entanglement entropy, the Hawking-Page phase transitions, etc, which could be seen directly from these two examples that we discussed here. 

In the next section we further inspect these connections using phase diagrams.

\section{Hawking-Page phase diagrams and thermodynamical ensembles}

To get a more unified picture of the connections between algebra and bulk reconstruction, we compare the Hawking-Page phase diagrams of BTZ and warped BTZ black holes in two theories of topologically massive gravity which is a chirally broken theory and new massive gravity which is a chirally symmetric theory, and in two different thermodynamical ensembles, where they have been constructed first in \cite{Detournay:2015ysa}. 

\begin{figure}[ht!]
\includegraphics[width=0.27\textwidth]{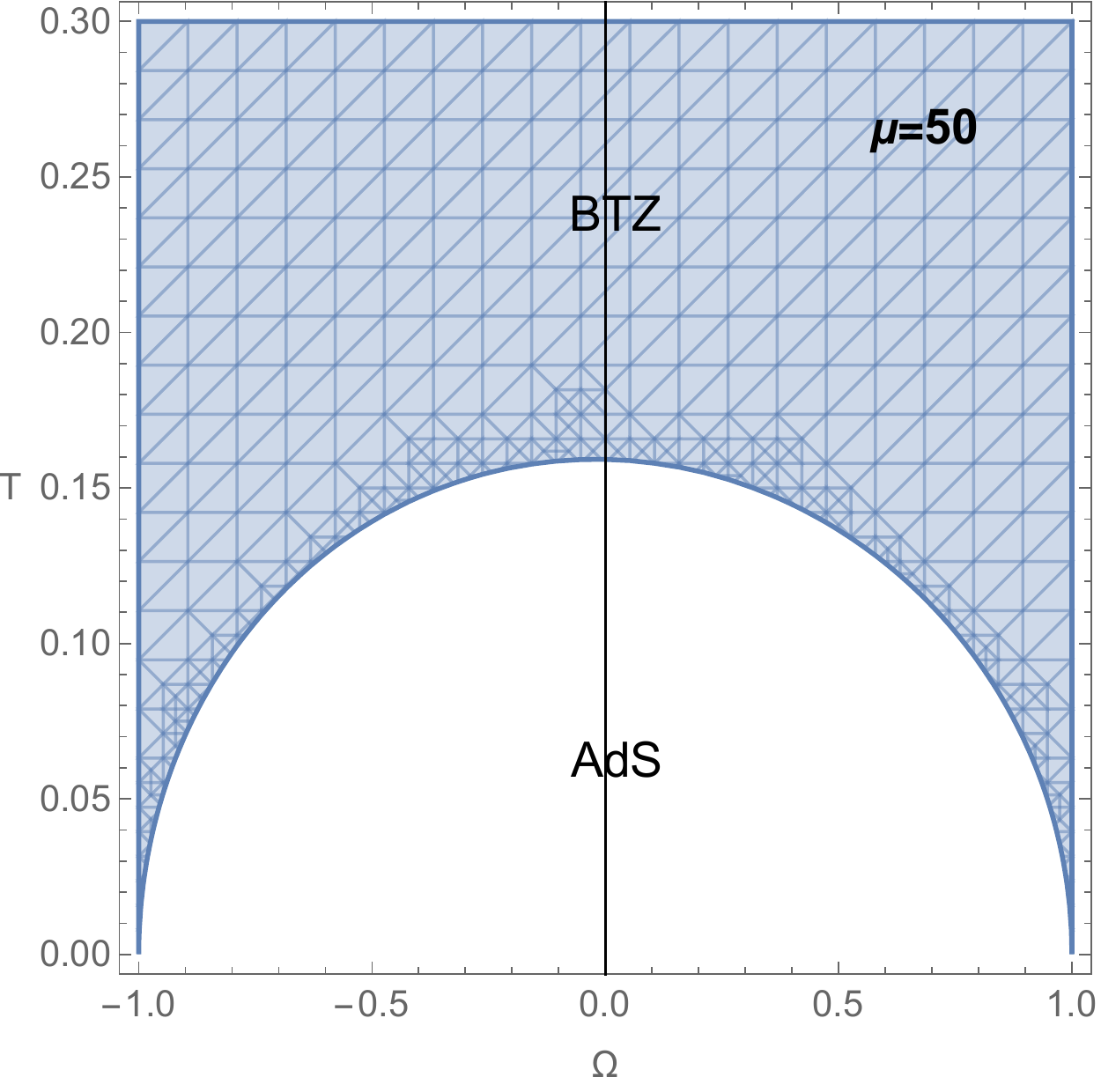} % Here is how to import EPS art
\includegraphics[width=0.27\textwidth]{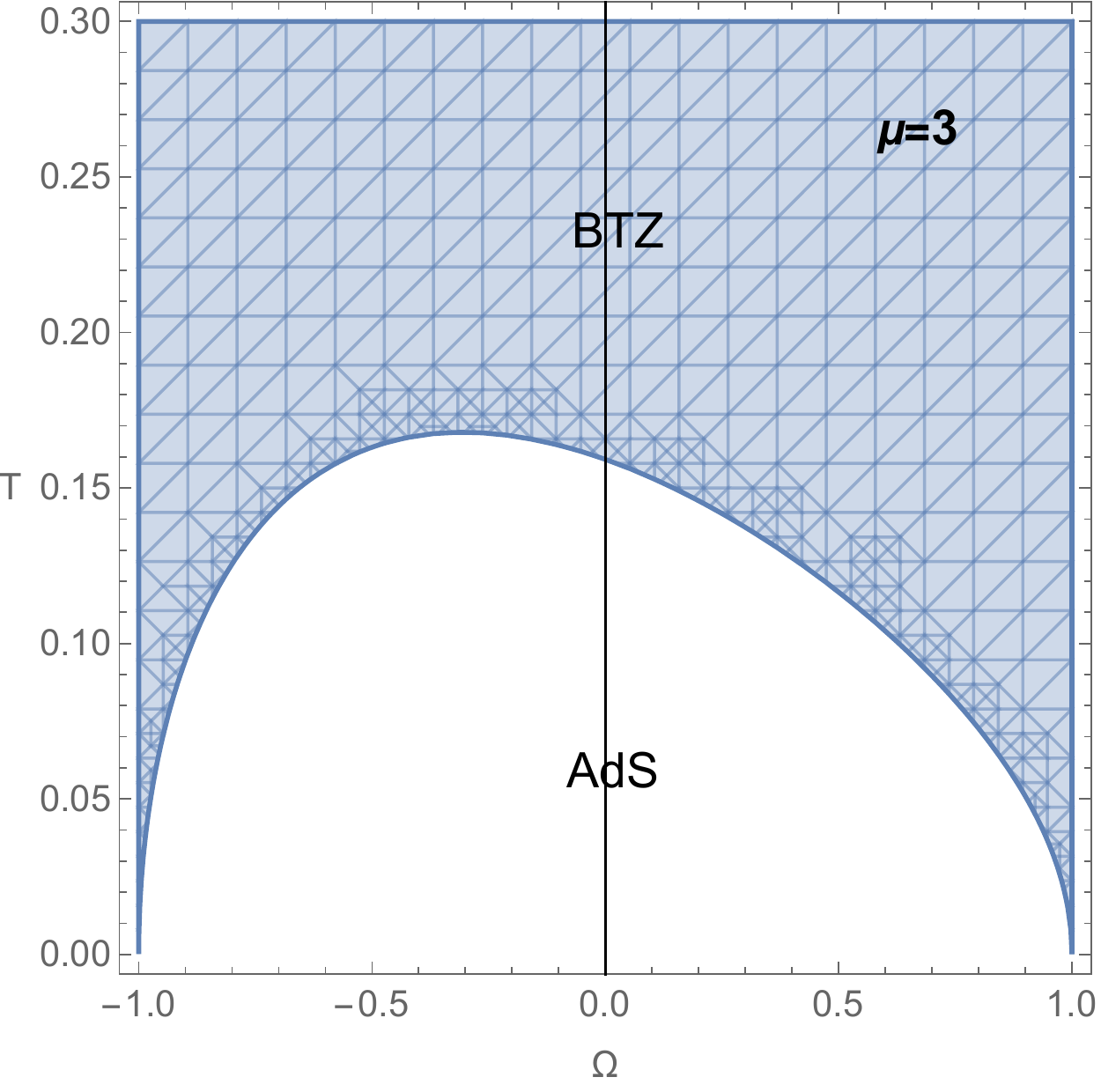} % Here is how to import EPS art
\includegraphics[width=0.27\textwidth]{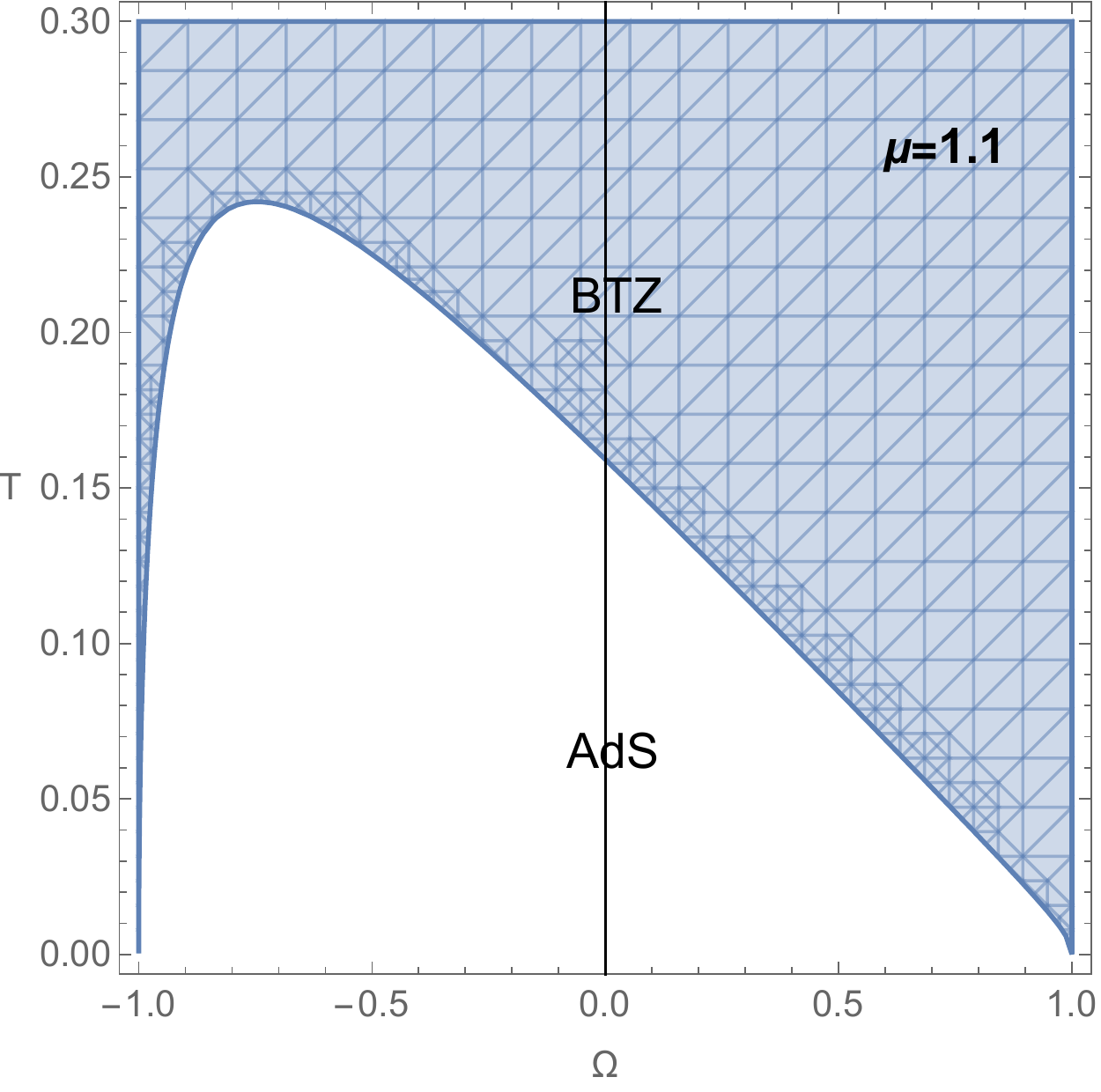} % Here is how to import EPS art 
\nonumber\\
\includegraphics[width=0.27\textwidth]{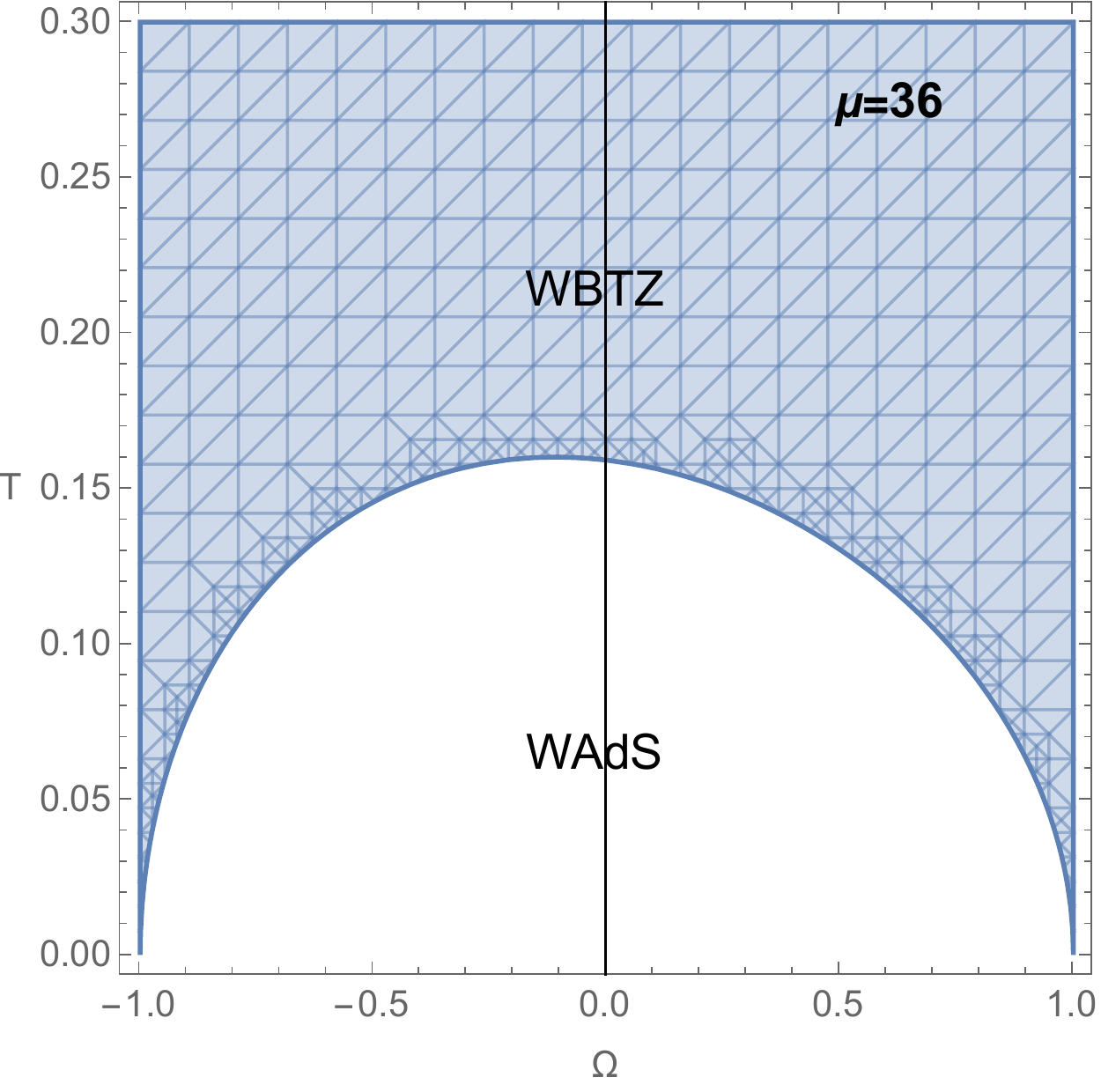} % Here is how to import EPS art
\includegraphics[width=0.27\textwidth]{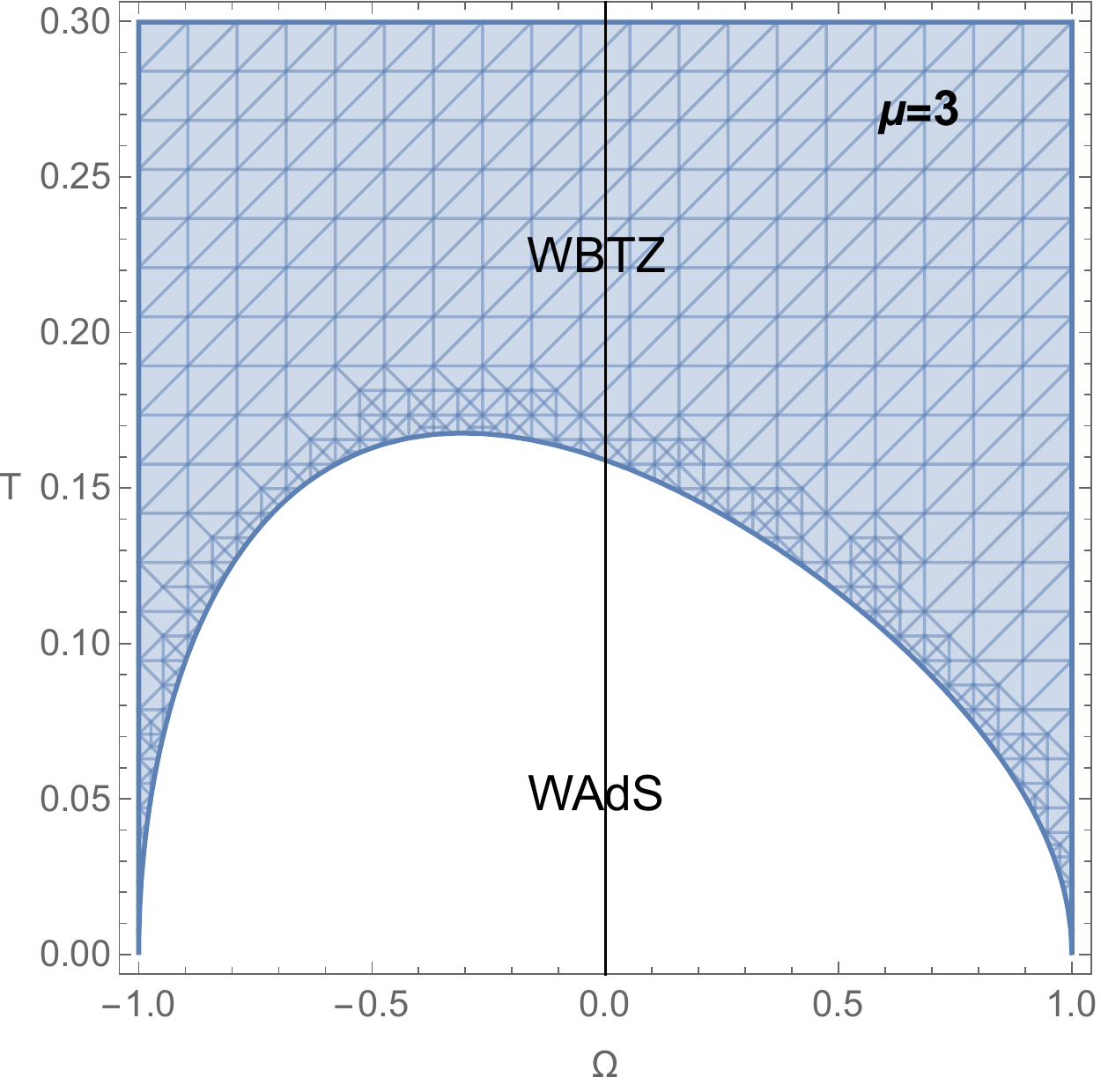} % Here is how to import EPS art
\includegraphics[width=0.27\textwidth]{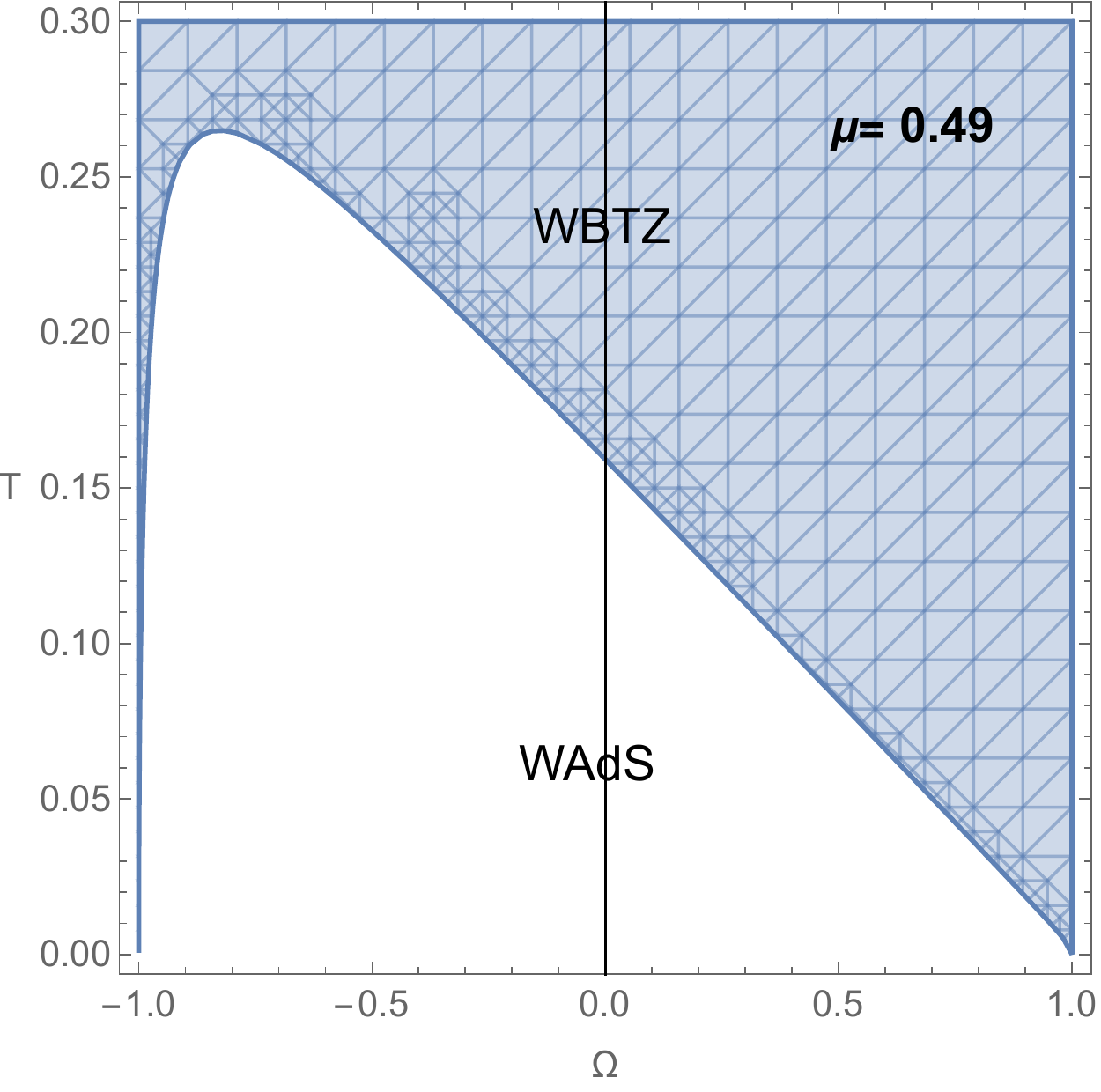} % Here is how to import EPS art
\caption{\label{fig:HPBTZ} The phase diagram of BTZ and warped BTZ black holes in the \textbf{chirally broken topologically massive gravity} and in the ``grand canonical ensemble'', where increasing the Chern-Simons parameter making the chirality to break further, would deform the diagrams more \cite{Ghodrati:2019bzz, Detournay:2015ysa}. Permission is granted to reuse the figures here.}
\end{figure}

\begin{figure}[ht!]
\includegraphics[width=0.27\textwidth]{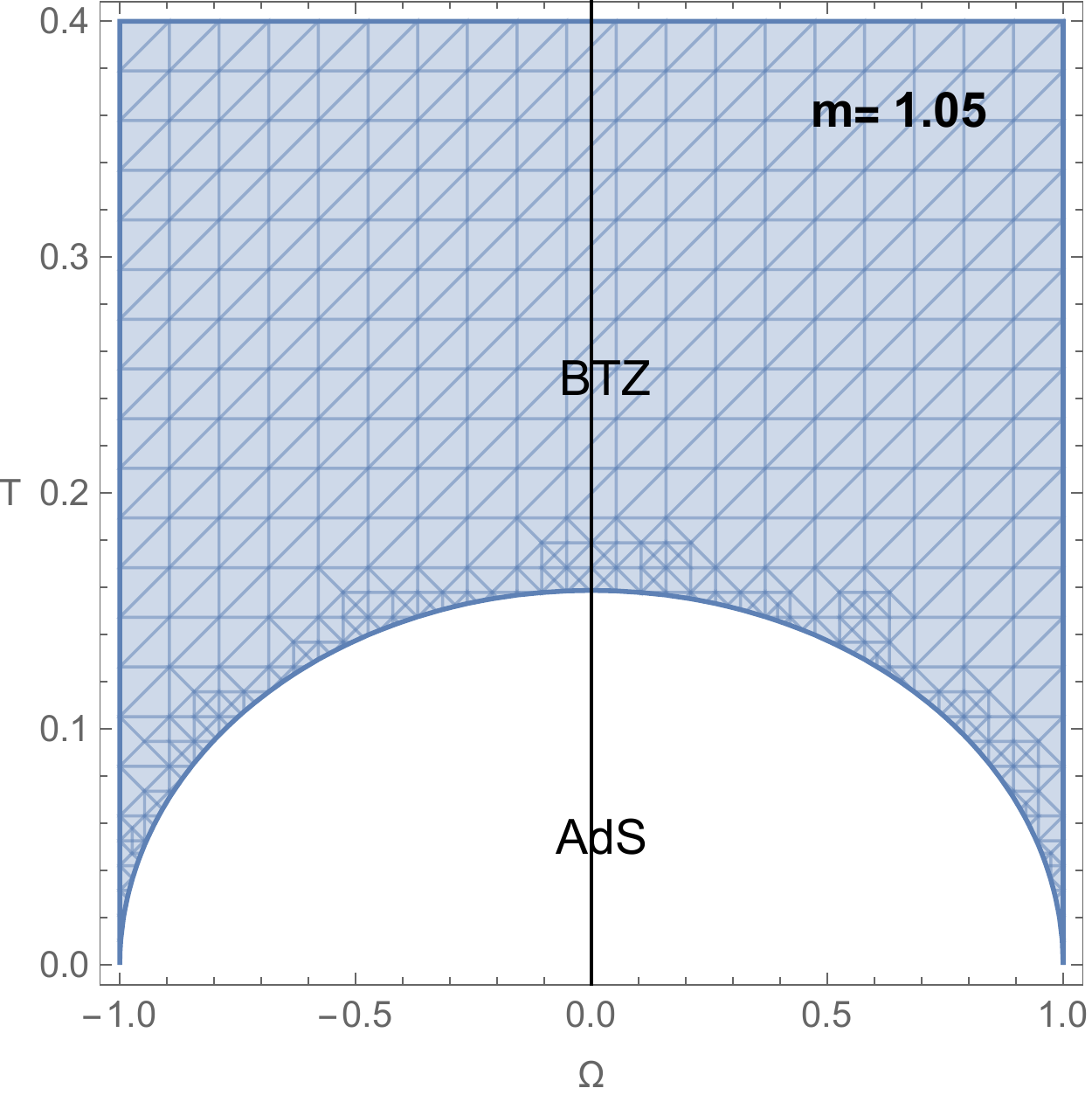}  \ \ \ \ \ \ \  
\includegraphics[width=0.27\textwidth]{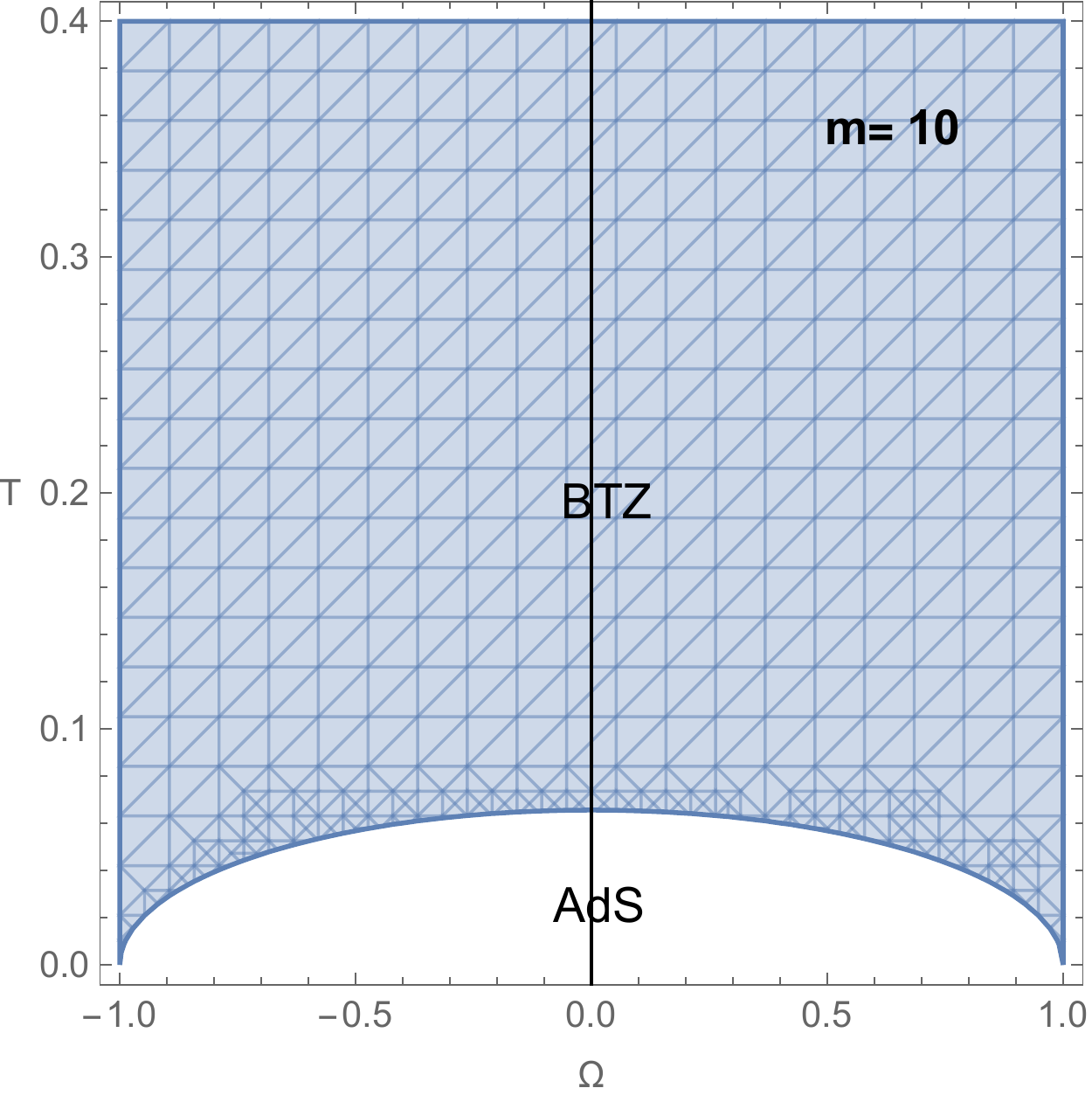} \ \ \ \ \ \ \  
\includegraphics[width=0.27\textwidth]{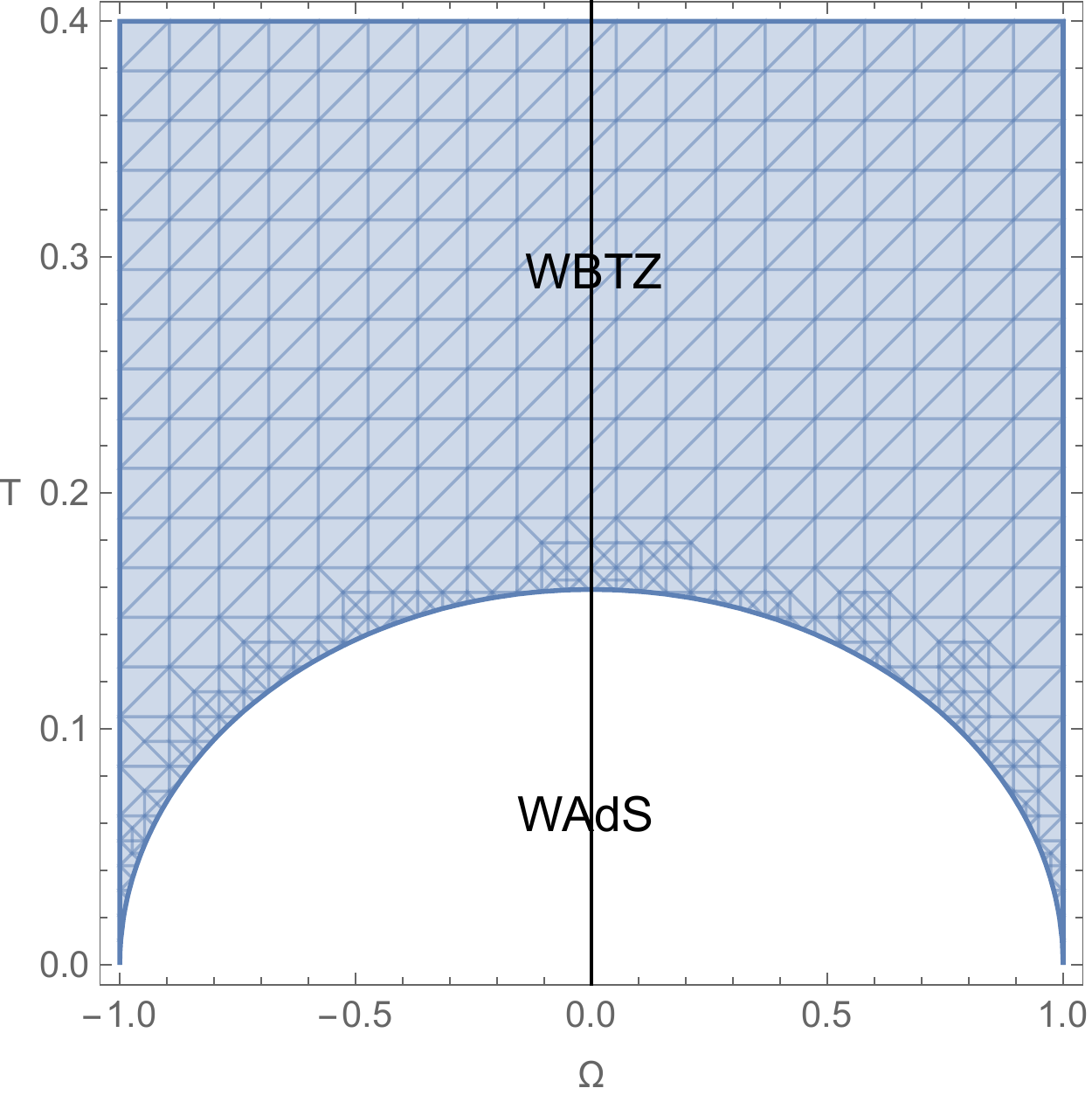}
\caption{\label{fig:NMGPBTZ} The phase diagrams of BTZ (for $m=1.05$ in the left and $m=10$ in the center) and warped BTZ black holes (for $c=l=1$ in the right part) in \textbf{chirally symmetric ``new massive gravity''} in the ``grand canonical ensemble'' \cite{Detournay:2015ysa, Ghodrati:2019bzz}. Permission is granted to reuse the figures here.}
\end{figure}

\begin{figure}[ht!]
\includegraphics[width=0.28\textwidth]{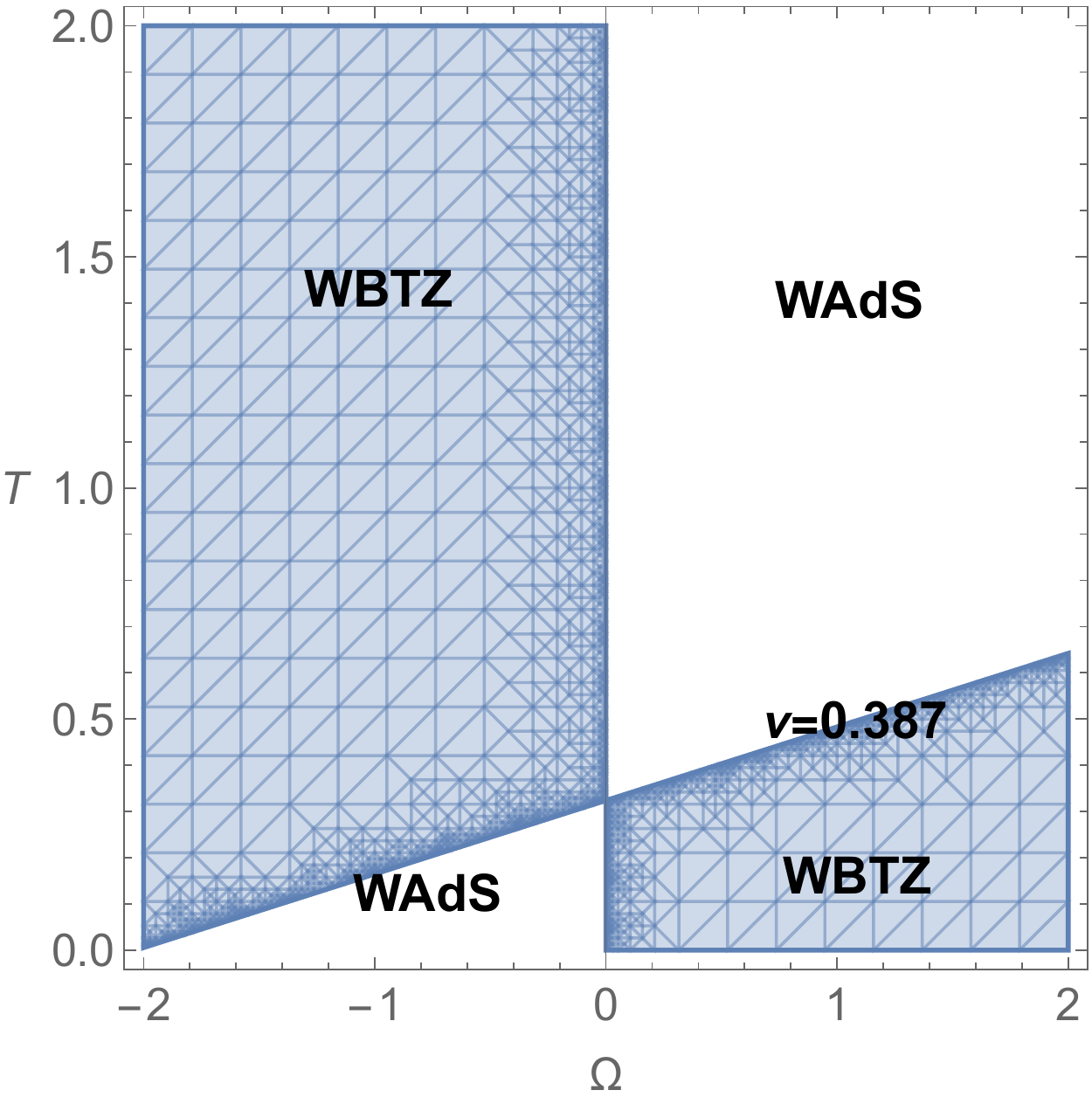} \ \ \ \ \ \ \ \  
\includegraphics[width=0.28\textwidth]{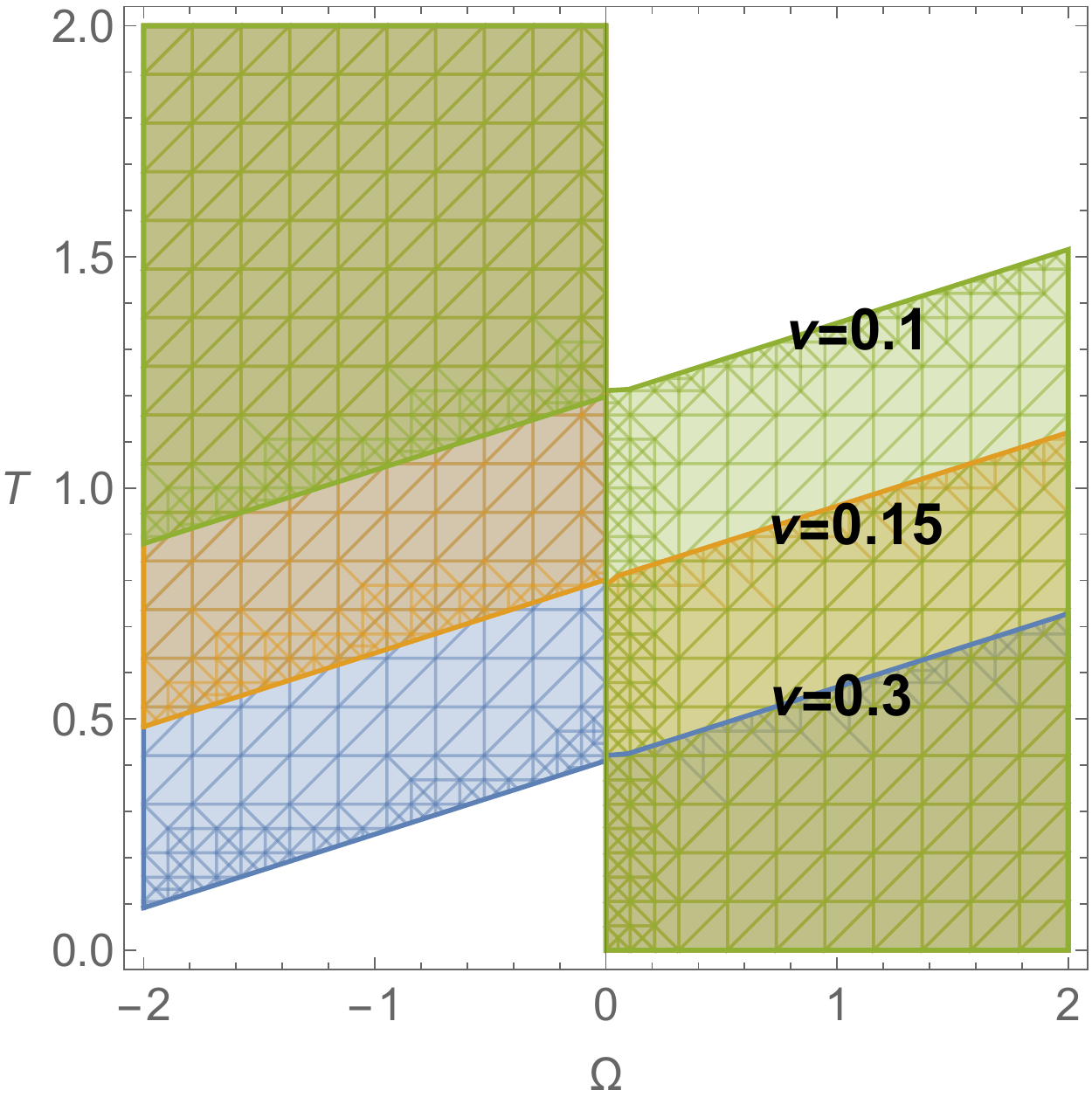}
\caption{\label{fig:nonlocal} The phase diagrams of BTZ and warped BTZ black holes in chirally symmetric  ``new massive gravity'' and in the ``non-local/quadratic canonical ensemble'' \cite{Ghodrati:2019bzz}.}
\end{figure}

The phase diagrams in the topologically massive gravity are shown in figure \ref{fig:HPBTZ}, and the phase diagrams in the new massive gravity are shown in figure \ref{fig:NMGPBTZ}. Both of these cases are constructed using the \textbf{grand canonical ensemble} which is actually the correct physical ensemble to study this theory. On the other hand, using another ensemble for the warped AdS solution, dubbed the "non-local/quadratic" ensemble, again the phase diagrams have been constructed as shown in figure \ref{fig:nonlocal}. 
These results demonstrate two points. First, the same solution in two different gravitational theories, would create two completely different phase structures, as for instance, the chirality of the theory can break the symmetries in the phase diagrams. Hence, one would expect the algebra of modes and the chaos bounds for the left or right moving modes depend on the structures of the gravity theory as well. Secondly, the choice of the thermodynamical ensemble is very important and one should use the "local ensembles" to study the chaos bounds and phase diagrams, and therefore the bulk reconstructions.

\section{Curvature invariants of Kerr (rotating) black holes }
In \cite{Abdelqader:2014vaa}, the curvature invariants of rotating BTZ black holes have been constructed and a new dimensionless parameter which specifies the local "Kerness" of the spacetime has been introduced.  In that work the curvature invariants specifically were used to study the physical properties of the spacetime around the "rotating" black holes. Considering the ability of curvature invariants to probe such properties, in addition to the recent works in modular chaos bound \cite{DeBoer:2019kdj}, and also the construction of Riemann curvature from the modular Berry phase  \cite{Czech:2019vih}, lead us in \cite{Ghodrati:2020mtx} to propose the connections between the modular scrambling modes in the boundary side which saturate the chaos bound, and the curvature invariants in the bulk gravity side. 

The bound for the modular chaos found in \cite{DeBoer:2019kdj} could be written as
\begin{gather}
\text{lim} \Big | \frac{d}{ds} \log F_{ij}(s) \Big | \le 2\pi,
\end{gather}
where
\begin{gather}
F_{ij}(s) = \Big | \langle \chi_i  | e^{i H_{\text{mod}}  s}  \delta H_{\text{mod} }  e^{-i H_{\text{mod}} s}  | \chi_j \rangle \Big |, \ \ \ \  \forall \ket{\chi_i} \in \mathcal{H}_{\text{code}}^\psi,
\end{gather}
are the matrix elements of the modular Hamiltonian.

The operators that saturate the bound are dubbed modular scrambling modes which could be written in the form of \cite{DeBoer:2019kdj}
\begin{gather}
G_+= \frac{1}{2\pi} \lim_{s\to\Lambda} e^{-2 \pi s} e^{-i H_{\text{mod}} s} \delta H_{\text{mod}} e^{i H_{\text{mod}} s}, \nonumber\\
G_-= - \frac{1}{2\pi} \lim_{s\to - \Lambda} e^{2 \pi s} e^{-i H_{\text{mod}} s} \delta H_{\text{mod}} e^{i H_{\text{mod}} s}.
\end{gather}

These modes satisfy the following relation \cite{DeBoer:2019kdj}
\begin{gather}
G_{\pm} \ket{\psi} \approx 0,
\end{gather}
and also as found in \cite{DeBoer:2019kdj}, they are the approximate isometries of the metric at the surface, since they satisfy the relation
\begin{gather}
\mathcal{L}_{\zeta_{\pm}} g_{\mu \nu} \Big |_{\text{RT}} \approx 0.
\end{gather}

Actually, the modular scrambling modes could be written using the curvature structures of the bulk metric, and their commutators also satisfy the relation  \cite{DeBoer:2019kdj}
\begin{gather}
\lbrack G_+ , G_- \rbrack = J^i (y) \partial_i + \delta x^+ \delta x^- R_{+ -} |^\mu _ \alpha x^\alpha \partial_\mu\nonumber\\
+ \left( \frac{1}{2} \epsilon_{-+} \nabla^i \delta x^- \nabla_i \delta x^+ - 2 J^i (y) a_i(y) \right ) (x^+ \partial_+ - x^- \partial_- ), 
\end{gather}
where $j^i = \frac{1}{2} ( \delta x^- \nabla ^i \delta x^+ - \delta x^+ \nabla^i \delta x^- )$.
 
 So considering these equations, we proposed in \cite{Ghodrati:2020mtx} that the resulting relation coming from applying the modular scrambling modes would in fact correspond to the syzygies and the curvature invariants in the bulk for each geometry, and thus another connection between the modular algebra and bulk curvature has been represented.

\section{Conclusion}

In this work, we clarified several subtleties in the connections between the boundary CFT data and the dual holographic bulk geometries. First, we showed that similar to rotating BTZ black holes, for the warped CFT/warped BTZ case, depending on the tilt parameter, two effective inverse temperatures could be defined, where each govern the behavior of $U(1)$ versus the $SL(2,R)$ modes, and accordingly the chaos in each direction, the tensor network structures, the entanglement entropy and the bulk reconstruction procedure in these theories. Second, we showed that it is very important to choose the correct thermodynamical ensemble to study chaos modes and phase transitions in various black holes, as the non-local ensembles lead to non-physical phase diagrams. Specifically, in the warped AdS case, the grand canonical ensemble instead of the quadratic non-local ensemble should be used. This is an important lesson in understanding the holographic bulk reconstruction as well.
Finally, we showed that the curvature invariants and syzygies in the bulk are related to the modular scrambling modes in the boundary CFT side, thusly pointing to another direct connection between the chaos algebra and bulk geometry.

\nocite{*}
\bibliography{aipsamp}% Produces the bibliography via BibTeX.

\end{document}